# Survey and comparison for Open and closed sources in cloud computing

Nadir K.Salih, Tianyi Zang
School of Computer Science and Engineering, Harbin Institute of Technology, China

***Abstract***

***Cloud computing is a new technology widely studied in recent years. Now there are many cloud platforms both in industry and in academic circle. How to understand and use these platforms is a big issue. A detailed comparison has been presented in this paper focused on the aspects such as the architecture, characteristics, application and so on. To know the differences between open source and close source in cloud environment we mention some examples for Software-as-a-Service, Platform-as-a-Service, and Infrastructure-as-a-Service. We made comparison between them. Before conclusion we demonstrate some convergences and differences between open and closed platform, but we realized open source should be the best option.***

## I. INTRODUCTION

Cloud Computing is a broad term that describes a many services. As with other significant developments in technology, many vendors have seized the term "Cloud" and are using it for products that sit outside of the common definition. In order to truly understand how the Cloud can be of value to an organization, it is first important to understand what the Cloud really is and its different components. Since the Cloud is a broad collection of services, organizations can choose where, when, and how they use Cloud Computing. In this paper we explain the different types of Cloud Computing services commonly referred to as Software as a Service (SaaS), Platform as a Service (PaaS) and Infrastructure as a Service (IaaS) and give some examples to compare between open and closed sources.

SaaS (software-as-a-Service) [22][34] provided to the consumer that use the provider's applications running on a cloud environment. The applications are accessible from various client devices through a web browser (e.g., web-based, Gmail, hotmail). The consumer does not manage or control the underlying cloud infrastructure including hardware or software resources see fig 1.

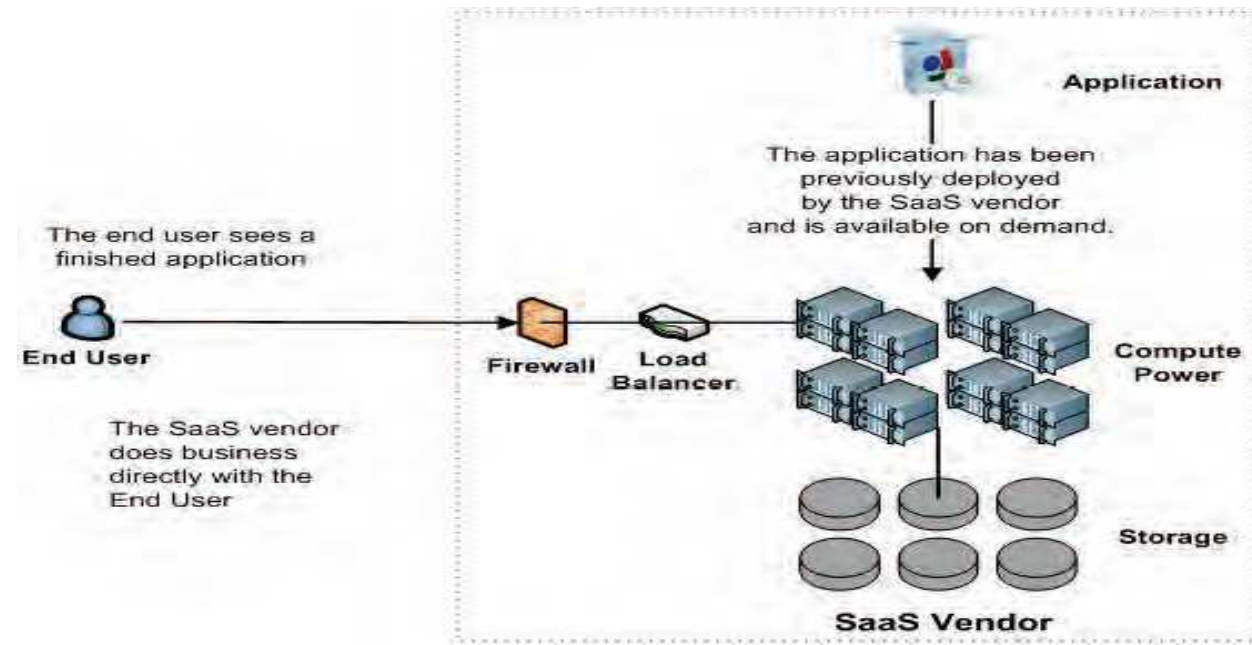

Fig 1 Software as a Service (SaaS)

SaaS have some features different from legacy application concept, like [23]:

- Access with web services.
- Administrator manager software.
- Software delivered in a "one to many model
- Users not required a lot of resources.
- Software can be integrated in one place.

PaaS (Platform-as-a-Service) is a form of cloud computing that help developer to optimize application [1]. PaaS platform has several advantages:(1) develop, test, deploy and maintain on the same integrated environment, which reduced development and maintenance costs; (2) users can seamlessly experience the software online without downloading or installing;(3) more closely integrated other online services and data; (4) built-in scalability, reliability and security;(5) improved the developer's cooperation; (6) in-depth understanding of user activity;(7) pricing based on actual usage. Current platforms tend to have its own feature, and PaaS vendors are trying to perfect their platform. Platform as a Service (PaaS) is the delivery of the computing platform, it reduces the complexity of managing underlying hardware and software layers and provides the facilities to support the complete life cycle of a web application [2][7]. PaaS provides a solution for offering multiple applications on the same platform thus increasing the economy of scale and reducing complexity [3].

Providers reduce risk in terms of upgrade cost of underlying platforms and allow Cloud users to concentrate on the application development. See Fig 2.

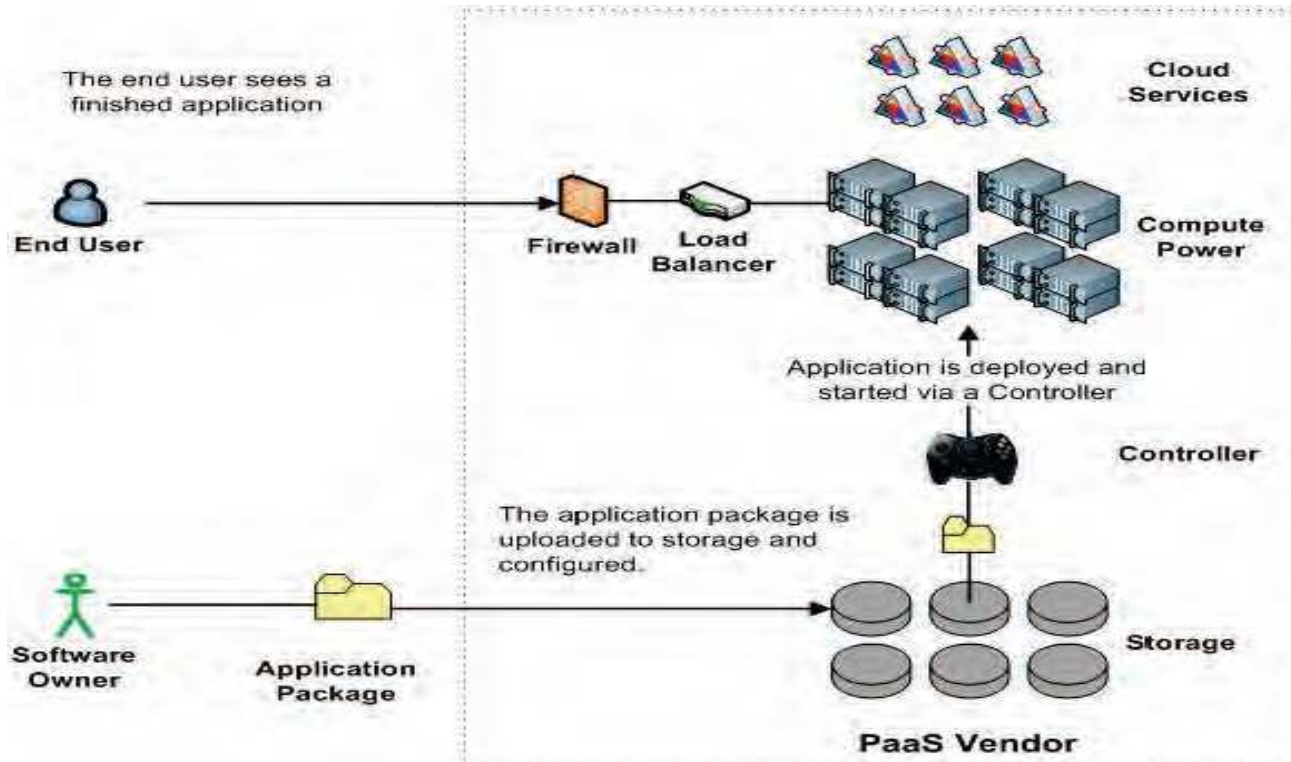

Fig 2 platform-as-a-Services

IaaS (Infrastructure-as-a-Service) see fig 3 referred to as Resource Clouds, provide (managed and scalable) resources as services to the user – in other words, they basically provide enhanced virtualization capabilities. According different resources may be provided via a service interface: Data & Storage Clouds deal with reliable access to data of potentially dynamic size, weighing resource usage with access requirements and / or quality definition. We can mention Characteristics of infrastructure as a service as [24]:

- Delivery of all resources as a service.
- Lower total cost of ownership.
- Full scalability.
- Eliminate the need for administration

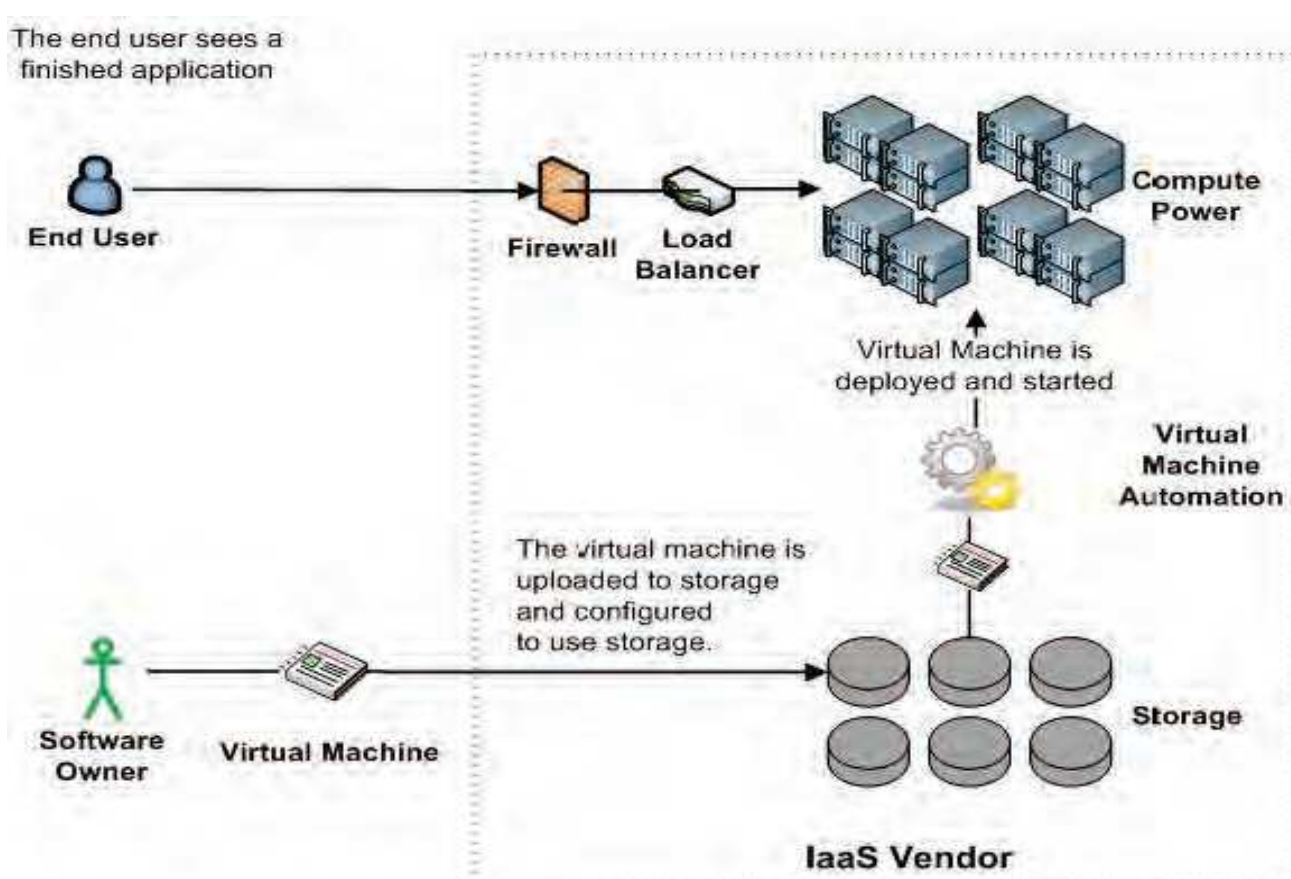

Fig 3 Infrastructure-as-a-Service

The remainder of this paper is organized as follows. After an overview of the cloud computing services is given in section1, and then the types of software-as-a-service are discussed in section 2. The some examples of the platform-as-a-service are mentioned in Section 3, Followed by kinds of infrastructure-as-a-service in section 4. Convergences of open and closed source in cloud computing are highlighted in Section 5. Finally, conclusions and future work are presented in Section 6.

## II. TYPES OF SOFTWARE AS A SERVICE

There are a several kinds of application or software defined as open or closed source for example:

*A. Autonomy Interwoven Team site CMS*

Interwoven [25] is owner to Interwoven TeamSite to leader in content management solutions. Interweaver's software and services enable organizations to effectively leverage content to drive business development by improving the customer working, increasing collaboration, and business processes in dynamic environments. Interwoven helps organizations extend and protect their sides, optimize their online presence, and provide a consistent and more engaging experience across all customer touch points. It enables organizations increasing productivity, simplifying business processes across various environments.

*B. AxCMS.net*

AxCMS.net [26] it is closed source and Content Management System (CMS) based entirely on Microsoft .NET platform that simplifies complex processes involved in creating and managing high scalable and interactive web system AxCMS.net guarantees brand and content consistency of multiple sites and in different languages while allowing employees and external resources to create, manage and publish web content. It helps you manage information and systems for your employees, partners and customers at any time. Increase productivity by making use of many employees.

*C. Contegro*

Contegro [27] it provided as commercial for design websites for big or a small company. It is very easy to use not need professional user.

*D. Amilia CMS*

Software Amilia [28] is CMS will not be installed directly on the web server, but in the amilia data center because it is proprietary. The requirements of the domain web server are very low. amilia only requires php. Amilia as SaaS brings with it many advantages. Amelia's functionalities and user interface based on the JavaScript DHTMLX framework and a sophisticated database management. The package comes in two different configurations: On Demand and Enterprise. The On Demand configuration is a (SaaS) model, can be accessed from any place. The new edition is deployed in the company network and can also be used with secure Internet access.

*E. Liferay Community Edition*

Liferay [29] as open source system provides a perfect web interface for many sources. It has interface called portal it self-contained interactive elements that written in very good way.

*F. AdaptCMS Lite*

AdaptCMS [30] it is free software and a content management system (CMS) written in PHP for front end, using a MySQL as backend. With AdaptCMS you can manage any website with an advanced option, custom fields, easily editable templates, an advanced permissions area and more that makes AdaptCMS suitable for any website.

*G. mojoPortal*

MojoPortal [31] [32] is an open source, look like (CMS) for ASP.NET which is written in high-level language. It supports plug-in and has built-in support for, among others application and an e-commerce feature. The project was awarded an Open Source Content Management System Award by Packt in 2007 saying that the "ease of use, set of relevant tools and plugins and also the fact that it is cross platform, made it stand out above the rest".

*H. Bricolage*

Bricolage is free and open source software. It is a content management system (CMS) written in the Perl programming language .it Designed to manage workflow for large websites with many contributors [33].

We can conclude the comparison between closed and open software-as-a-service at the follow:

1- Software is protected by copyrights and is sold to users to earn money, while in open source available free.

2- To develop the software is easy and quickly in open source rather than closed

3- The successful issue in quality of closed source is related to centralized management while in open source the quality is from its openness because many programmers examine it and can detect bugs.

4- Security terms in open source are much better than closed source.

5- Terms of documentation and central management, there is a problem with open source rather than closed source. Due to lack of responsible person for the projects, the users should wait until the problem has been resolved in anyway.

There are some comparisons between closed and open source of software-as-a-service that we mentioned see table 1 in appendix.

## III. TYPES OF PLATFORM-AS-A-SERVICE

There are several solutions available in the PaaS market, to mention a few:

*I. Windows Azure*

The Windows Azure Platform [8], [6] support a lot of applications and used many language PHP, Java and C. work well with databases .it owner to Microsoft.

*J. Google App Engine*

From a big provider Google has close platform called Google App Engine [9], [4] it used web service in full utilization. The important feature of it exactly designs for real time dynamic application.

*K. Force.com*

Another closed platform that is running business application and used WSDL it called Force.com [10] platform .It is very easy to apply any user or organization their idea inside application. It treated database modification in different tables as events occur.

*L. Manjrasoft Aneka*

New proprietary platform but it is spread widely to support multiple programming models .It called Aneka [5] using in various scientific researches and commercial environment.

*M. Red Hat OpenShift*

Red Hat OpenShift [11] support for Java EE it is open source leadership .It can encourage developers to build and management their application according to their idea without constraints

*N. VMware*

VMware [12] it can be used as hybrid cloud platform furthermore it comes to use for wide area of application framework and development. It portable approach to delivering new applications has emerged.

*O. TioLive*

TioLive [13] the free Source Platform as a Service (PaaS) and ERP/CRM Software as a Service (SaaS), arrived the release of TioLive Grid. With TioLive Grid, everyone can now run on his own servers a Private Cloud or a hybrid Cloud at no license cost. TioLive Grid aims at giving more freedom to user and offers total control over important business data. It is the first step towards Distributed Cloud Computing, a new approach to Cloud Computing which will replace on the next year's legacy Centralized Cloud Computing solutions controlled by proprietary SaaS vendors

*P. WSO2 Stratos*

WSO2 [14] is open source cloud platform for enterprise applications. WSO2 Stratos offers organizations of all sizes a fully hosted application (PaaS). Using WSO2 Stratos, IT professionals can build and used applications and services with instant provisioning of enterprise servers, including the portal, Enterprise Service Bus (ESB), and application server. WSO2 Stratos is exist today as an early adopter release for private clouds, as a develop version on public clouds, and as an early release of the downloadable open source software. As a fully open source solution WSO2 Stratos does not require any licensing fees.

Platform as a Service (PaaS) offers the potential to democratize web development by enabling anyone who can use a browser to assemble and extend web-based applications. Yet early PaaS players have introduced PaaS solutions that are remarkably owner, introducing high switching costs to move data or logic from one PaaS provider to another. In contrast, an Open-source Platform as a Service (OPaaS) solution leverages industry standards and allows applications to be deployed across multiple cloud providers. An OPaaS solution has four characteristics:

1. Open source solution is available as open source and supported by a large open source community.
2. Portable**,** developers can deploy OPaaS applications on multiple cloud infrastructures, including public and private clouds.
3. Open server platform – all developers will be able to use standard languages and existing code within the OPaaS.
4. Flexible client platform - developers and end users must have standard, easy-to-use tools for configuring OPaaS user interface.

Although this features and other likes good security and low cost in open source platform, but it didn't developed like closed platform. See table 2 in appendix compared these two types of platform.

## VI. TYPES INFRASTRUCTURE-AS-A-SERVICE

For infrastructure-as-a-Service there are a lot of closed and open types. We can take some examples like:

*Q. Agathon Group*

Agathon Group [15] it is closed source intended to enhance the use of the Internet by preventing unacceptable use. Some users of Agathon Group's Internet services those who access some of their Services but do not have accounts not like who pay a monthly service fee to subscribe to the Services.

*R. Amazon EC2*

Amazon Web Services (AWS) [16] it is not free but it has huge usage in elastic platform. It is very easy to use and only pay for what is used.

*S. Cisco*

Cisco's [17] it is closed but suitable for public cloud network called the inter-cloud. It includes a long-term market transition define by ubiquitous portable workloads and a rich cloud environment in which external and internal clouds share resources. The inter-cloud will allow secure and resources management to help developer.

*T. IBM*

IBM Company has not free example of IaaS [18]. It can capable for manage all resources in addition it prefer cloud by add virtualization issues.

*U. Eucalyptus*

The Eucalyptus it works in IaaS as open source [19] [20] [21]. It used virtual machine in controlling and manages of resources. It not only easy to use but has compatible with other provider like Amazon EC2.

*V. Open Nebula*

OpenNebula it has integration with various environments [20] [21].It can be work through command line interface and web service. In addition it applies adminstarion for resources through virtual machine.

*W. Reservoir*

It is open source and defines as reference architecture for next generation of IaaS. It has a lot of features like automating the services provisioning and scalability, open source code, and virtualization technology independent.

*X. Nimbus*

Nimbus [21] it is very good open source for IaaS work in administration of virtual network. It supported by Secure Shell SSH into all compute nodes.

Cloud computing performances depend on different parameters such as the CPU speed, the amount of memory, network and hard drive speed. In virtual environment the hardware is shared between virtual machines. And open source have some features like:

1- Datacenter operators can easily build cloud services within their existing infrastructure to offer on-demand, elastic cloud services.

2- Open source IaaS software platform, which enables users to build, manage and deploy compute cloud environments.

There are some characteristics as the same in closed source see table 3 in appendix.

## V. CONVERGENCES AND DIFFERENCES

Some cloud service providers use open-source software or platforms, the base systems are usually proprietary. However, there are a few entirely open-source based platforms, as well as applications and tools available to manage mainly IaaS cloud services. These tools allow the user to monitor, manage and control the virtual instances. Unfortunately, most open-sources are at the IaaS or PaaS level and very few SaaS open-source applications exist. We found from aforementioned comparison some convergences between open and closed source like:

1- The languages supported can be the same in two sources

2- Application support and compatibility look like the same.

3- Using of web services in two types

4- Interoperability of operating system.

5- Other features like monitoring

Although this convergences but the important thing stand against closed system it is the services costing. In addition the security issue can the best in open source because the control with user or organization. For that the future will be own the open source type.

## VI. CONCLUSIONS

This paper compares the two closed source development model and open source model for cloud computing. The result shows that in terms of costing is better to use open source methodology, rather than closed source, while there is a problem with documentation and design of using open source model, because of using novice volunteers and make a useless documentation. Also we have found out what the convergences and differences of closed source software and open source.

In future we want to make a model that used open source in three layers of service in cloud computing SaaS, PaaS, and IaaS .To be full control by user and measure performance and quality of services.

## ACKNOWLEDGMENT

This work has been developed with the support under the project with number: 2012AA02A604, 863 Program key projects in China: The Technology and the System Development for Smart Acquirement of Personal Healthcare Information. And so the Key Project of NSF in China: Methodology of Value-oriented Software Services: Theory, Method and Application with number: 61033005.

## Appendix

| | Autonomy Interwoven Teamsite CMS | AxCMS.net | Contegro | amilia CMS | Liferay Community Edition | AdaptCMS Lite | mojoPortal | Bricolage |
|---|---|---|---|---|---|---|---|---|
| Platform | Perl, Java | ASP.NET | ASP.NET | PHP | Java | PHP | ASP.NET | Perl on mod_perl |
| Supported | Oracle, SQL Server, DB2 | SQL Server | SQL Server | MySQL | HSQLDB, MySQL, Oracle, SQL Server, DB2, Apache Derby, Informix, InterBase, JDataStore | MySQL | SQL Server, MySQL, PostgreSQL, SQLite, Firebird, SQLCE | MySQL, PostgreSQL |
| Web management | Yes | Yes | Yes | Yes | yes | yes | yes | yes |
| Software provider | Interwoven TeamSite | Axinom | Kiwi CMS | Amilia Corporation Inc | Liferay .com | Charlie Page's | Packt | Salon.com, |
| Software type | Closed | Closed | Closed | Closed | Open | Open | Open | Open |

Table1 comparison of open and closed source in Software-as-a-Service

| | Windows Azure | Google App Engine | Force.com | Manjrasoft Aneka | Red Hat OpenShift | VMware | TioLive | WSO2 Stratos |
|---|---|---|---|---|---|---|---|---|
| Service type | Web and none web application | Web app | Web services | Compute/data ,Web and non-web apps | Web app | Simplified infrastructures | Web app | SOA middleware services |
| OS support | Windows | Windows or Linux | Apex | Linux, Windows | Linux | Linux, Windows | GNU/Linux | Linux, Window |
| Deployment language | Visual Studio, and .Net C#, C++, | Python, java | Apex | Java | java EE | PHP,java | java | PHP,java |
| User access interface | Microsoft windows azure portal | Web-based administration console | Adobe's Flex | Workbench, web-based portal | Command line. GUI. | vSphere Web | ERP5 | Command line. GUI |
| Compatibility | with the Microsoft app | Amazon's EC2, S3 | Amazon's EC2 | Amazon's EC2, Xen | Amazon EC2 | Eucalyptus, Amazon's EC2 | Amazon EC2, Gpars | Amazon's EC2, Eucalyptus, Ubuntu Enterprise Cloud |
| Source type | closed | Closed | Closed | Closed | Open | Open | Open | Open |
| Owner | Microsoft | Google | salesforce.com | Manjrasoft | Red Hat | VMware Inc | Nexedi | WSO2 |

Table2 comparison of open and closed source in platform-as-a-Service

| | Agathon Group | Amazon EC2 | Cisco | IBM | Eucalyptus | OpenNebula | Reservoir | Nimbus |
|---|---|---|---|---|---|---|---|---|
| Provider | Agathon Group | Amazon EC2 | Cisco | IBM | Eucalyptus | OpenNebula | Reservoir | Melia Technologies |
| OS | Gentoo,Linux,Windows server 2008 | Gentoo,Linux,Windows | CentOS, ,Linux, Windows server 2008 | Red Hat Enterprise Linux SUSE Linux | Cent OS | Linux,Open solaris,open SUSE | Linux | Microsoft Windows XP/Vista |
| Language Supported | PHP, XMLmosaic | Java,php,python,ruby, WinDev | - | all the programming languages | Java | Java, Perl, PHP, | Java, Perl, PHP,SAS,SQL | HTML |
| Monitoring | No | Yes free | Yes free | Yes | Yes, Free | Yes, Free | Yes, Free | yes |
| Web Service | Yes, Free | No | Yes, Free | Yes | No | No | Yes, Free | yes |
| Control Interface | API (Application Programming Interface) | API (Application Programming Interface),command line | Web Based Application/ Control Panel | Web Based Application/Control Panel,API | Web Based Application/ Control Panel | Web Based Application/ Control Panel | API (Application Programming Interface) | Web Services based |
| Source type | Closed | Closed | Closed | Closed | Open | Open | Open | Open |

Table3 comparison of open and closed source in Infrastructure-as-a-Service